\documentclass[manuscript]{aastex}

\newcommand{\myemail}{mtakuma@kwasan.kyoto-u.ac.jp}

\shorttitle{Temporal power spectra of the horizontal velocity of the solar photosphere}
\shortauthors{\textsc{MATSUMOTO} and \textsc{KITAI}}

\begin{document}

\title{Temporal power spectra of the horizontal velocity of the solar photosphere}

\author{Takuma \textsc{MATSUMOTO} and Reizaburo \textsc{KITAI}}
\affil{Kwasan and Hida Observatories, Kyoto University
   , Yamashina-ku, Kyoto, 607-8417, Japan; \myemail}
   
\begin{abstract}
We have derived the temporal power spectra of the horizontal velocity of the solar photosphere. The data sets for 14 quiet regions observed with the \textit{G}-band filter of \textit{Hinode}/SOT are analyzed to measure the temporal fluctuation of the horizontal velocity by using the local correlation 
tracking (LCT) method.
Among the high resolution ($\sim$0\farcs2) and seeing-free data sets of \textit{Hinode}/SOT, we selected the observations whose duration is longer than 70 minutes and cadence is about 30 s.
The so-called $k$-$\omega$ diagrams of the photospheric horizontal velocity are derived for the 
first time to investigate the temporal evolution of convection.
The power spectra derived from $k$-$\omega$ diagrams typically have a double power law shape bent over at a frequency of 4.7 mHz.
The power law index in the high frequency range is -2.4 while the power law index in the low frequency range is -0.6.
The root mean square of the horizontal speed is about 1.1 km s$^{-1}$ when we use a tracer size of 0\farcs4 in LCT method.
Autocorrelation functions of intensity fluctuation, horizontal velocity, and its spatial derivatives 
are also derived in order to measure the correlation time of the stochastic photospheric motion.
Since one of possible energy sources of the coronal heating is the photospheric convection, the power spectra derived in the present study will be of high value to quantitatively justify various coronal heating models.
\end{abstract}

\section{INTRODUCTION}

At least three types of convection are known to ubiquitously exist at the solar surface: 
granules, mesogranules, and supergranules.
Their properties such as size, lifetimes, shapes have been studied by many authors 
(see reviews by, e.g., Leighton 1963, Spruit et al. 1990, Nordlund et al. 2009 and references therein).
Through the magnetic field, the convective energy is transported upward to supply 
the sufficient energy to maintain the 1 MK corona.
Therefore, the dynamics of the convective motions plays a key role in the so-called coronal heating problem, 
one of the most important issues in the solar physics. 
However, the temporal evolution of convection motion have not been sufficiently elucidated.


The vertical motion in convection is easily observed by using Doppler analysis. This motion 
generates compressible waves such as magnetohydrodynamic slow mode waves or fast mode waves.
Although a lot of energy can be transported upward by the waves caused by convection,
compressible waves are not considered to contribute to coronal heating, because the 
wave energy will be significantly reduced before reaching the corona by 
shock dissipation or reflection in the chromosphere \citep{holl81}.
On the other hand, the horizontal motion in convection plays an important role in the coronal heating.
The interaction between magnetic field and the convection generates Alfv\'{e}n waves \citep{uchi74} while
the stochastic photospheric motion braids the field lines to store the energy in the corona as electric current \citep{park83}.
Due to the lack of observations, it has been poorly understood what mechanisms contribute to coronal heating.

The local correlation tracking (LCT) method commonly is used to derive the horizontal velocity field \citep{nove88,berg98}.
Since LCT method uses apparent motion of granules to derive the velocity, it is better to 
use high spatial resolution and seeing-free data sets.
From the horizontal velocity derived by LCT method, mesogranules and supergranules 
can be observed (e.g. Kitai et al. 1997, Ueno \& Kitai 1998, Shine et al. 2000).
Although various studies reveal the frequency distribution \citep{titl89,berg98} and 
the spatial power spectra \citep{rieu00, rieu08} of horizontal velocity fields, 
few studies explicitly mention the temporal power spectra \citep{tarb90}.
In a previous study, we derived the photospheric horizontal velocity in 14 different quiet regions 
to justify the Alfv\'{e}n wave model for coronal heating \citep{mats10}.
In this study, we will show the temporal evolution of the horizontal motion of the photosphere in more detail.
When discerning between various coronal heating models (see reviews by, e.g.,  Mandrini et al. 2000, 
Aschwanden et al. 2001, Klimchuk 2006, 
and references therein), the power spectra derived in the present study will be of high value.

\section{OBSERVATION AND DATA REDUCTION}

\textit{Hinode} was launched on September 22, 2006 in order to investigate the unsolved problems of 
solar physics, such as coronal heating.
Solar Optical Telescope (SOT) \citep{tsun08,suem08,ichi08,shim08}, one of the three telescopes mounted on \textit{Hinode}, 
is an optical telescope whose spatial resolution is unprecedentedly high (0\farcs2-0\farcs3, or 150-200 km) 
for a solar telescope in space.
As opposed to ground based observations that are affected by atmospheric seeing, 
seeing-free data sets over a long time span can be continuously obtained by SOT.
We selected 14 data sets of continuously observed quiet regions with the \textit{G}-band filter
between October 31, 2006 and December 29, 2007.
Our basic selection criteria was for the data set to have a duration longer than 70 minutes, with a mean time cadence less than 32 s. 
Duration of the selected data sets ranges from 75 min to 345 min and the cadence is almost constant with the accuracy of 1 second.
Our data sets are selected to be located within $\pm$ 100 arcsec from the disk center so that the line of sight is within 6 degrees from the local normal.
Pixel resolution of the CCD is 0\farcs054 (40 km) and FOV is larger than 27\arcsec$\times$ 27\arcsec  (20,000 km $\times$ 20,000 km).

We applied dark current subtraction and flat-fielding in the standard manner for all of the \textit{G}-band images.
The solar rotation, tracking error, and satellite jitter causes overall displacements between two images.
Using cross correlation between two consecutive images, we have derived the optimum 
shift to adjust the overall displacements in an accuracy of sub pixel scale.

Besides granular motion, $p$ mode waves generally disturb the intensity in the \textit{G}-band movies.
In order to reduce the contribution from $p$ modes, we have applied so called sub-sonic filtering technique \citep{titl89} to our data set.
The sub-sonic filtering technique removes the power where $\omega / k > V_{ph}$ in Fourier space.
In the present study, we have selected $V_{ph} = 5$ km s$^{-1}$.

\begin{figure}
\epsscale{1.0}
\plotone{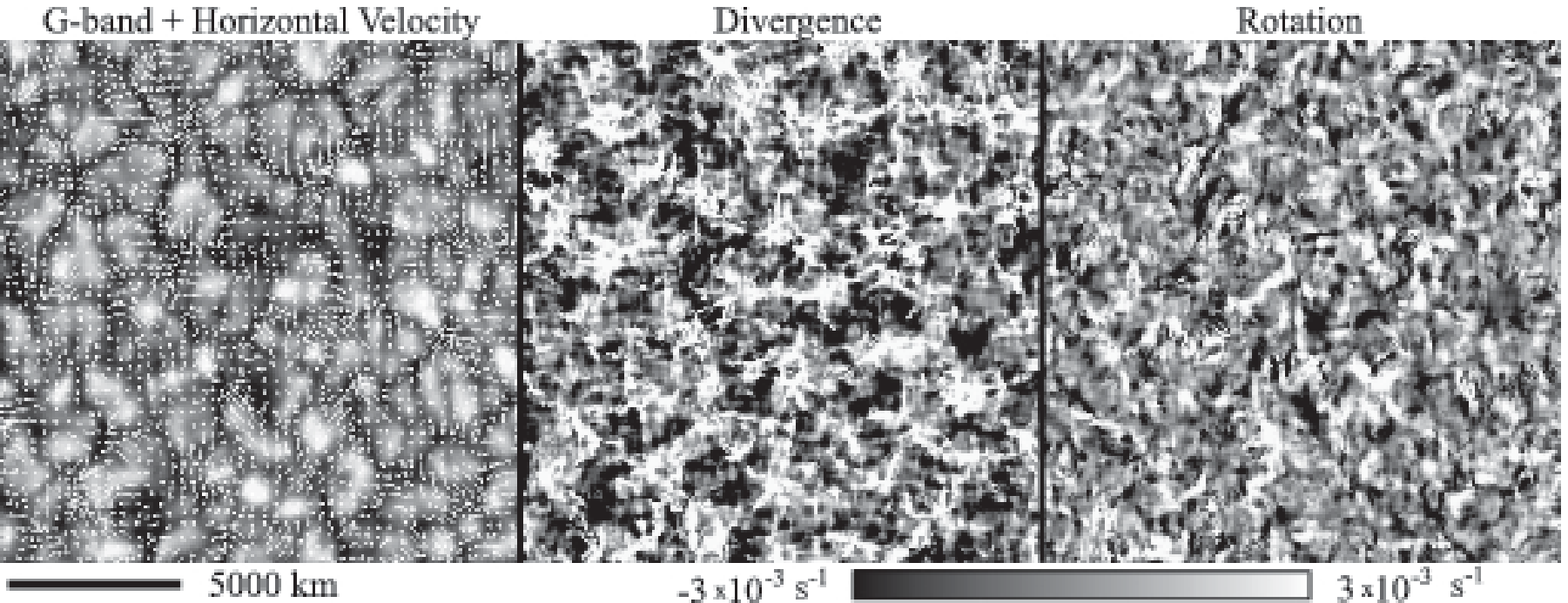}
\caption{
\textit{G}-band image with horizontal velocity arrows (left), divergence (middle), and rotation (right).
The velocity field and its derivatives are averaged over $10^3$ s.
 \label{LCTfield}}
\end{figure}

The photospheric horizontal velocity of granular motion can be obtained by applying the LCT technique to the \textit{G}-band time series.
In the LCT techniques developed by \cite{berg98}, rectangular subfields or ``tiles" are used, and the displacement of tiles between two consecutive images determines the horizontal velocity. 
Although the amplitude of the LCT velocity usually depends on the size of the tracer, we have fixed the tile size to be 0\farcs4 in the present study, the same resolution as in the study of \cite{berg98}.
From the horizontal velocity fields, divergence ($\nabla \cdot V = \partial _x V_x + \partial _y V_y$) 
and rotation ($\nabla \times V |_z = \partial_x V_y- \partial _y V_x$) can be derived, 
where $x,y$ are the coordinates in the photospheric plane and $z$ is the vertical coordinate.
Figure \ref{LCTfield} shows the \textit{G}-band intensity image with horizontal velocity arrows (left), 
divergence (middle), and rotation (right), averaged over $10^3$ s.


\section{RESULTS AND DISCUSSIONS}

\begin{figure}
\epsscale{1.0}
\plotone{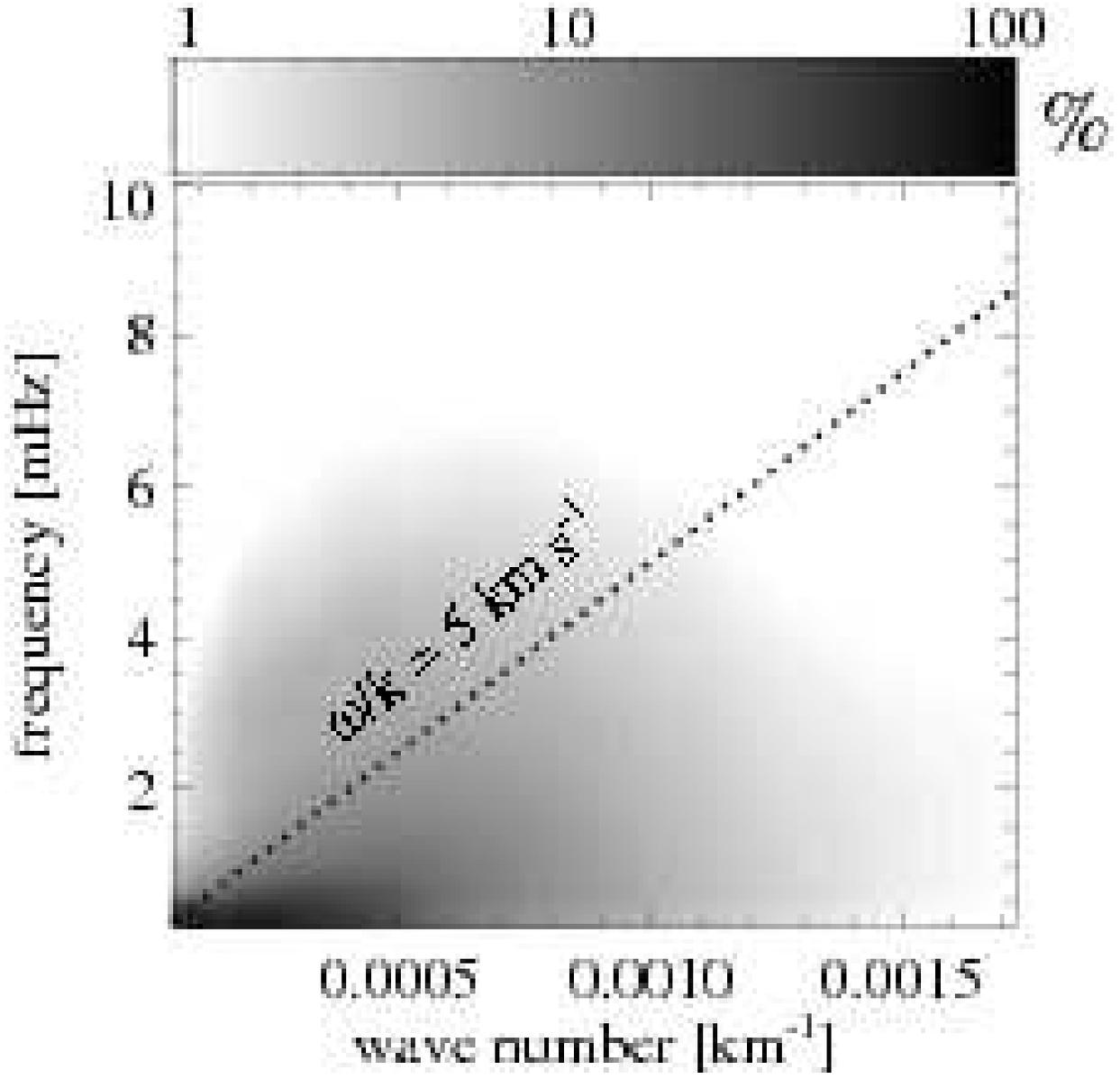}
\caption{
$k$-$\omega$ diagram of horizontal velocity fluctuation. The dotted line indicates $\omega=$ 5 km s$^{-1}~k$. The
gray scale color represents the power spectrum density normalized by the maximum power in log scale. \label{kwdiagram}}
\end{figure}

We derived the apparent velocity in \textit{G}-band data from \textit{Hinode}/SOT by using the LCT method.
Figure \ref{kwdiagram} shows the so-called $k$-$\omega$ diagram of the horizontal velocity
 normalized by the maximum power.
The time series of LCT flow maps were Fourier transformed to estimate the power spectra.
 The dotted line in the figure represents $\omega / k=5~\mathrm{km~s}^{-1}$.
Since we use the subsonic filtering method, spectral power above the dotted line is significantly reduced.
The region in the lower frequency and the smaller wave number has greater power.

\begin{figure}
\epsscale{1.0}
\plotone{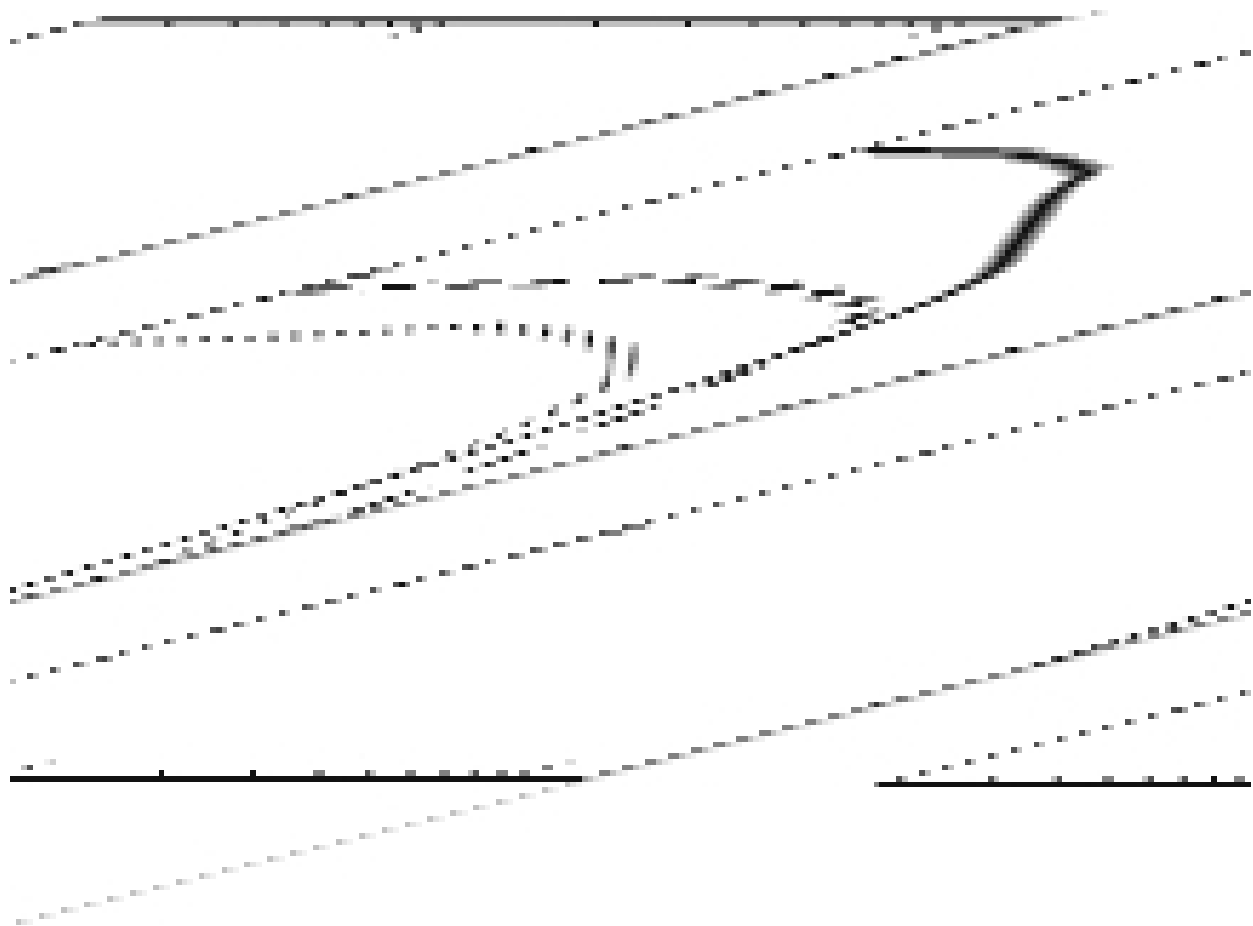}
\caption{
  Temporal power spectrum density of the photospheric horizontal velocity (solid line), rotation (dotted line), and divergence (dashed line).
  After the horizontal velocity is derived by the LCT technique, the power spectrum is estimated by FFT.
 \label{tspectrum}}
\end{figure}

Integrating $k$-$\omega$ diagram over wave number space, 
we can estimate the frequency power spectrum density, $P_{\nu}$.
The power spectrum density is defined as, 
	\begin{equation}
		<V_{\perp}^2> = \int _{\nu_{min}} ^{\nu_{max}}~P_{\nu}~d\nu \label{eq_power},
	\end{equation}
where $V_{\perp}$ is the LCT velocity, $\nu_{min}$ is the lowest frequency determined by total duration, and $\nu_{max}$ is the highest frequency coming from the observational sampling time. 
The symbol, $<>$, denotes the temporal average. 
Figure \ref{tspectrum} shows an example of the power spectrum density.
Power spectra can be fitted to double power-law function, $P_\nu \propto \nu^{\alpha_L}$ when $\nu < \nu_b$ (so-called break frequency) and $P_\nu \propto \nu^{\alpha_H}$ when $\nu \ge \nu_b$. 
Typically, $\sqrt{<V_{\perp}>},\nu_b, \alpha_L, $ and $ \alpha_H$ of LCT velocity are 1.1 km s$^{-1}$, 4.7 mHz, -0.6, and -2.4, respectively in this work.
By using the same analysis, temporal power spectra of divergence and rotation of velocity fields are also estimated (the dashed line and the dotted line in figure \ref{tspectrum}). 
The total power, $\alpha_L$, and $\alpha_H$ of the divergence is 8.4 $\times$ 10$^{-3}$ s$^{-1}$, -0.2, -1.5 respectively and the divergence will contribute to produce the acoustic mode waves.
The total power, $\alpha_L$, and $\alpha_H$ of the rotation is 6.8 $\times$ 10$^{-3}$ s$^{-1}$, -0.2, -1.3 respectively and the rotation is related to the generation of torsional Alfv\'{e}n waves.
The power spectra of the LCT velocity has a harder power law index than 
that of its derivatives ($\nabla \cdot V, \nabla \times V$).


The resulting power spectra are quite essential for evaluating models of coronal heating by Alfv\'{e}n waves \citep{holl82,kudo99,suzu05,suzu06,mats10}.
In their numerical simulations, nonlinear Alfv\'{e}n waves are driven by photospheric convection to heat the corona. \cite{mats10} include the observed temporal power spectra shown here to generate Alfv\'{e}n waves and succeed in producing sufficient energy flux to heat the corona. 


Next, in order to derive the``correlation time" of the stochastic convection motion, 
we derived the autocorrelation function (ACF), 
\begin{eqnarray}
	R(\tau) = {\left< f(t) f(t+\tau) \right> \over <f(t)^2>}\label{eq_acf}
\end{eqnarray}
, where $f$ is one of physical variables associated with turbulent motion and $<>$ denotes 
temporal averaging process.
The correlation time is determined as the e-folding time of the ACF. 
Correlation time often scales with coronal heating rate in various heating models that use 
stochastic convection motion \citep{stur81, park88, gals96}.
Figure \ref{crtime} shows the ACF of the \textit{G}-band intensity fluctuation (solid line), horizontal velocity (dashed line), divergence (dotted line), and rotation (dash-dotted line).
For \textit{G}-band intensity, the correlation time is about 250 sec, which is comparable with the value in \cite{titl89} 
and references therein. The horizontal velocity has shorter correlation time ($\sim$ 100 sec) than the \textit{G}-band intensity fluctuation. 
\cite{rieu00} also derived the ACF of the horizontal velocity and found a relatively larger correlation time of about 35-45 min, because they were looking at larger spatial/temporal scale phenomena. 
ACFs of rotation and divergence fall rapidly and their correlation times are about 50 sec.
The correlation time in the coronal heating models is usually defined by the velocity correlation time 
(e.g. Sturrock \& Uchida 1981).
However, the lifetime of granules, which corresponds to the intensity correlation time,
is usually used as the correlation time in the previous studies.
Since the velocity correlation time turned out to be shorter than the intensity correlation time, 
the coronal heating rate will be greatly reduced.

Let us explain why the correlation time of intensity fluctuation ($\tau _I$) is longer than that of velocity ($\tau _V$) qualitatively.
The power spectrum of intensity ($P_I$) can be regarded as the power spectrum of spatial scale.
From the dimensional analysis, the power spectrum of velocity ($P_V$) is proportional to $P_I \nu ^2$ 
so that $P_I$ has a spectrum that is at most a power 2 softer than $P_V$.
Since Fourier transform of ACF corresponds to power spectrum, softening of the power spectrum 
means hardening of the ACF decay or long correlation time. 
Therefore it is reasonable that $\tau_I$ is longer than $\tau_V$. 
From the similar dimensional analysis, the shorter correlation time of divergence or rotation 
can be explained.

\begin{figure}
\epsscale{1.0}
\plotone{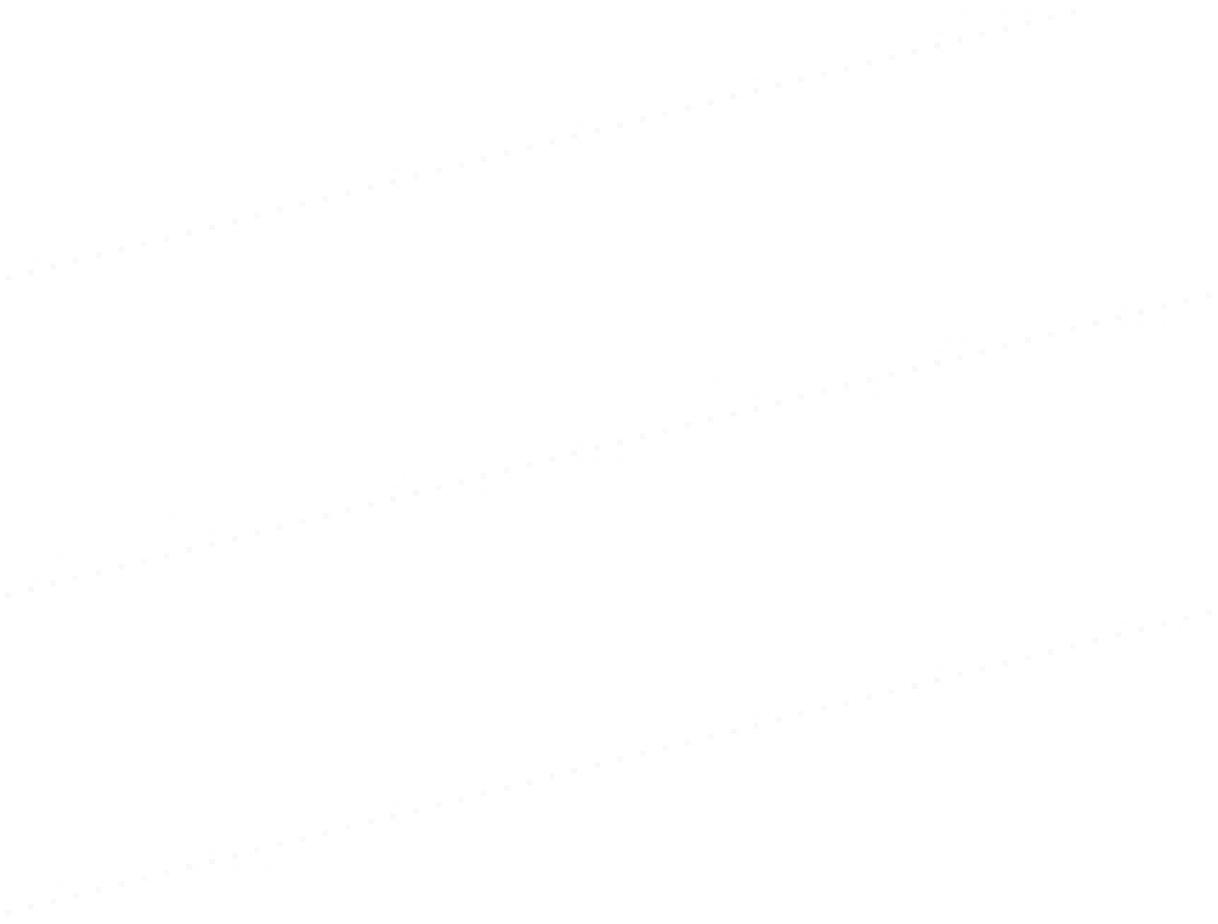}
\caption{
Spatial power spectrum density of the photospheric horizontal velocity.  The power law index of the 
dotted line is -5/3.
\label{turb}}
\end{figure}

If we integrate $k$-$\omega$ diagram over frequency space, the spatial power spectrum density can 
be estimated (Figure \ref{turb}). The power law index for the high $k$ range ($k \sim 10^{-3}$) becomes -5/3, 
which represents the Kolmogorov type turbulence.
Our results are complementary to the results of \cite{rieu08} since they investigate large spatial structures.
\cite{espa93} showed the same power law index in the power spectra of vertical photospheric velocity.
These results indicate that the granules are Kolmogorov turbulent eddies.

We also analyze motion of test particles in the LCT flow field. 
From the spatial deviation of test particles ($\delta$) and the elapsed time ($\tau$), diffusion coefficient ($D_{ph} \equiv \delta ^2 / 2 \tau$) can be derived, if we assume motion of each test particle can be approximated by random walk.
For our data set, the diffusion coefficient $D_{ph}$ is around 500 km$^2$ s$^{-1}$, which is comparable with 
that of \cite{tarb90}. When we apply this value to the electric current cascading coronal heating model \citep{ball86}, heating rate that scales with the diffusion coefficient of the photospheric horizontal motion becomes less than half of the required rate.
Therefore when we consider coronal heating in the quiet sun, current cascading may not contribute significantly to the heating rate.

Since the LCT method tracks the apparent motion of the photospheric convection, 
the velocity derived here would contain some artifacts from the apparent velocity even though we have 
carefully removed the effect of acoustic waves by subsonic filtering method.
The decrease in density or increase in temperature causes the destruction of CH molecules, 
resulting in \textit{G}-band intensity changes \citep{steiner01}.
These intensity changes can affect the LCT velocity.
It is important to compare the result of LCT measurements with the recent realistic 3D simulations
\citep{stei98,rieu01,geor06,geor07}.

In conclusion, $k$-$\omega$ diagrams of LCT velocity are derived for the first time to investigate 
the temporal evolution of the photospheric convection. 
Integrating the power spectra over wave number space generally reveals a double power law spectral shape whose break frequency is about 4.7 mHz. The power law index in the low frequency range is -0.6 while the power law index in the high frequency range is -2.4.
The correlation time of \textit{G}-band intensity and horizontal velocity is about 200 sec and 100 sec respectively while that of divergence and rotation is 50 sec.
The diffusion coefficient of photospheric convection is around 500 km$^2$ s$^{-1}$, which is not sufficient to heat the corona above quiet region through current cascading.
By using the power spectra derived in the present study, some of the coronal heating models will be justified (e.g. Matsumoto \& Shibata 2010).

\acknowledgments

\textit{Hinode} is a Japanese mission developed and launched by ISAS/JAXA, with NAOJ as domestic partner and NASA and STFC (UK) as international partners. 
It is operated by these agencies in co-operation with ESA and NSC (Norway).
Scientific operation of the \textit{Hinode} mission is conducted by the \textit{Hinode} science team organized at ISAS/JAXA. 
This team mainly consists of scientists from institutes in the partner countries. 
Support for the post-launch operation is provided by JAXA and NAOJ (Japan), STFC (U.K.), NASA, ESA, and NSC (Norway).
Takuma Matsumoto is supported by the Research Fellowship from the Japan Society for the Promotion of Science for Young Scientists.

\begin{figure}
\epsscale{1.0}
\plotone{f4.eps}
\caption{
Normalized autocorrelation function of \textit{G}-band intensity fluctuation (solid line), horizontal velocity (dotted line),  rotation (dashed line), and divergence (dash dotted line) defined by the equation \ref{eq_acf}.
 \label{crtime}}
\end{figure}

\clearpage


\end{document}